\documentclass[aps,prb,preprint,superscriptaddress]{revtex4-2}
\usepackage{amsmath}
\usepackage{amsfonts}
\usepackage{xcolor}
\usepackage{graphicx}
\usepackage{epsfig}

\begin{document}
	
	\title{Extended Hybridization Expansion Solver for Impurity Models with Retarded Interactions}
	
	
	\author{Lei Gu}
	\email[]{gulei@sicnu.edu.cn}
	%\thanks{}
	\affiliation{College of Physics and Electronic Engineering, Sichuan Normal University, Chengdu 610068, China}
	
	\author{Jia Luo}
	\affiliation{College of Physics and Electronic Engineering, Sichuan Normal University, Chengdu 610068, China}
	
	\author{Ruqian Wu}
	\affiliation{Department of Physics and Astronomy, University of California, Irvine, California 92697, USA}
	
	\author{Guoping Zhao}
	\email[]{zhaogp@uestc.edu.cn}
	%\thanks{}
	\affiliation{College of Physics and Electronic Engineering, Sichuan Normal University, Chengdu 610068, China}
	\affiliation{Center for Magnetism and Spintronics, Sichuan Normal University, Chengdu 610068, China}

	\begin{abstract}
       We extend the continuous-time hybridization expansion solver to a general form, where the hybridization function and retarded interaction are treated on equal footing. Correlation functions can now be directly obtained via functional derivatives with respect to the bosonic propagators, similar to the measurement of Green's functions. We devise a combinatorial scheme of measuring the correlation function, whose efficiency partially emulates that of the Green's function measurement. The algorithm and numerical methods are validated through application to an impurity model involving both electron-phonon coupling and exchange interactions, a case where the previous hybridization expansion algorithm is not applicable. Our improvement of the hybridization expansion solver promotes its applicability in studies of electron-phonon coupling, the extended dynamical mean field theory, and the dual boson method.
	\end{abstract}
	
	%\keywords{}
	
	\maketitle
	
	\section{Introduction}
	The continuous-time quantum Monte Carlo algorithms~\cite{CTMC_RMP2006,Prokof1998,Rombouts1999,Rubtsov2004,Rubtsov2005,Gull2008,Gull2008b,Gull2011,Werner2006,Werner2006b} constitute a class of solvers for impurity models and are extensively applied in the implementation of dynamical mean-field theory (DMFT)~\cite{DMFT_RMP1996,Metzner1989,Georges1992} and its extensions~\cite{DiagEDMFT_RMP2018,Hafermann2009,Rubtsov2009,Hafermann2010,Ayral2013,Ayral2015,Ayral2016,Iskakov2016,Stepanov2019,Brener2020,Matteo2022,Stepanov2022}. By expanding the partition function to terms with certain spans in imaginary time, the numerical schemes avoid explicit time discretization, thereby minimizing the risk of large discretizatoin errors. Among those continuous-time algorithms, the hybridization expansion (CT-HYB) method~\cite{Werner2006,Werner2006b} possesses additional advantages, including relatively small average expansion order near phase transitions~\cite{Gull2007} and versatility for multiorbital models with complex interactions~\cite{Stepanov2021,Matteo2022,Gukelberger2017,Zang2022,Rai2024}. Furthermore, various numerical techniques have been proposed to enhance its efficiency. The Monte Carlo sampling is implemented via a Markov chain in the configuration space, so partition of the Hilbert space~\cite{Haule2007,Lauchli2009} and skip-list based data management~\cite{Semon2014} can considerably facilitate calculating the rate of configuration proposals. The measurement procedure accommodates compact representations of Green's~\cite{Boehnke2011,Shinaoka2017,Hafermann2012,Shinaoka2018}, which substantially compress and denoise the results. When incorporated with the inchworm quantum Monte Carlo method~\cite{Cohen2015}, the possible sign problem can be notably mitigated~\cite{Eidelstein2020,Strand2024}.

	In the context of DMFT, retarded interactions arises from the interaction between electrons and bosons, typically coupling to phonons~\cite{Werner2007,Werner2010,Otsuki2013} or to an effective boson bath in the extended dynamical mean field theory (EDMFT)~\cite{Sengupta1995,Si1996,Smith2000,Chitra2000,Chitra2001,Sun2002,Huang2014} that represents the inter-site interactions. Integrating out the bosonic degrees of freedom leads to a retarded interaction. When the retarded interaction term commutes with the local Hamiltonian, the rate of a configuration is modified by a weight containing an interaction between all hybridization events ~\cite{CTMC_RMP2006,Werner2010}, which can be computed at essentially negligible computational cost~\cite{Hafermann2013}. However, limited cases, such as systems with density-density interactions, can satisfy this commutation requirement, making systems with inter-site exchange interactions  unsuitable. As a diagrammatic extension of EDMFT~\cite{DiagEDMFT_RMP2018}, the dual boson method~\cite{Rubtsov2012,vanLoon2014,Hafermann2014,vanLoon2014b,Stepanov2016,Stepanov2016b,Peters2019,Vandelli2020,Harkov2021} also involves solving an impurity model with retarded interactions. Currently, when applying the CT-HYB solver to systems with inter-site exchange interactions, the retarded interactions are neglected. In other words, the retarded interaction is not determined self-consistently, but calculated posteriorly after solving the impurity with hybridization functions only~\cite{Vandelli2020}. The development of a CT-HYB algorithm that can handle both hybridization and retarded interaction would significantly broaden its applicability.

	Among the continuous-time quantum impurity solvers, the interaction expansion method can be generalized to handle retarded interactions~\cite{Assaad2007}, where the original on-site interaction and the retarded interaction are treated on equal foot. On the other hand, because the expansion procedure of the partition function is generic, it is feasible to carry out a hybrid expansion within the CT-HYB framework, even in the presence of both the hybridization function and retarded interactions. Since density or spin operators are composed of a creation and an annihilation operator, they are of bosonic nature. The interchange between two of these operators, or with fermionic creation (annihilation) operators does not involve sign change, suggesting that a hybrid expansion scheme would not exacerbate the sign problem. Furthermore, as a retarded interaction term consists of two density (spin) operators linked by an bosonic propagator, the corresponding correlation function can be measured via functional derivative, similar to the Green's function measurement in prior CT-HYB.

	An ingenious idea underpinning the effectiveness of CT-HYB is to group all the configurations containing hybridization events at the same set of time points~\cite{Werner2006}. This procedure not only substantially suppresses the sign problem, but also enable efficient measurement of Green's functions, as the value for every pair of time points in the set can be measured at a single Monte Carlo sampling. Its efficient implementation relies on an elegant formulation of the ratio between minors and determinants of a matrix~\cite{Rubtsov2005,Boehnke2011,Park2011}. On the other hand, interchanging two bosonic operators does not lead to sign change, so applying the idea to retarded interactions results in a sum over all permutations, rather than a determinant. We find that a forceful grouping does not improve efficiency but instead complicates the implementation, as there is no analogous elegant ratio formulation in this case. However, we will demonstrate that, even without configuration grouping, the correlation function for each pair of time points in a configuration can still be measured through combinatorial operations. In other word, while the configuration sampling for retarded interactions is less efficient, the efficiency of correlation function measurement can match that of Green's function measurement.

	The paper is organized as follows: The partition function expansion, formulation of correlation function measurement, and the proposal probabilities for insertion and removal of a retarded interaction term are derived in Sec.~\ref{II}, where we further propose the combinatorial measurement scheme. In Sec.~\ref{III}, the algorithms are tested through an impurity model with inter-orbital exchange interaction and Holstein electron-phonon coupling, which violates the commutation requirement for the efficient handling of the retarded interaction. We demonstrate that the direct measurement of correlation functions via function derivative can be more reliable than computing them from two-particle Green's functions. We also show that the Legendre representation can significantly reduce noise in the correlation function measurement. Following the conclusion remarks, we provide an appendix discussing how time ordering cancel the $1/k!$ expansion factors.

	\section{Formulism and algorithm\label{II}}
	\subsection{Partition function expansion}
	The impurity model action with both hybridization functions and retarded interactions takes the form
	\begin{equation}
		\mathcal{S} = \mathcal{S}_{at} + \mathcal{S}_{hyb} + \mathcal{S}_{ret},\label{act}
	\end{equation} 
	where
	\begin{align}
	&\mathcal{S}_{at} = \int_{0}^{\beta}d\tau \sum_{\nu} \bar{c}_{\nu}(\tau)[\partial_{\tau}-\mu]c_{\nu}(\tau) + \int_{0}^{\beta}d\tau\mathcal{H}_{loc},\\
	&\mathcal{S}_{hyb} = -\int_{0}^{\beta}d\tau\int_{0}^{\beta}d\tau' \sum_{\nu\nu'} \bar{c}_{\nu}(\tau)\Delta_{\nu\nu'}(\tau-\tau')c_{\nu'}(\tau')\label{hyb}\\
	&\mathcal{S}_{ret} =-\int_{0}^{\beta}d\tau\int_{0}^{\beta}d\tau'\sum_{\sigma\sigma'}\phi_{\sigma}(\tau)D_{\sigma\sigma'}(\tau-\tau')\phi_{\sigma'}(\tau').\label{ret}
	\end{align} 
	Here, $\mathcal{S}_{at}$ is the atomic part of the action, and $\mathcal{H}_{loc}$ includes orbital potential energies and interactions within the impurity. $\mathcal{S}_{hyb}$ represents the hybridization, where $\bar{c}_{\nu}$ and $c_{\nu'}$ are Grassmann variables corresponding to operators $c^{\dagger}_{\nu}$ and $c_{\nu'}$, with $\nu$ and $\nu'$ indexing spin-orbital flavors of the impurity electrons. In the retarded interaction $\mathcal{S}_{ret}$, the bosonic variables $\phi_{\sigma}$ ($\phi_{\sigma'}$) correspond to composite operator in the form $c^{\dagger}_{\nu}c_{\nu'}$. Here $\mathcal{S}_{ret}$ is not written in an explicitly Hermitain form, and we treat the conjugate of $\phi_{\sigma}$ as a variable of a different flavor.
	
	The expansion of the partition function proceeds as follows:
	\begin{align}
		Z =& \int \mathcal{D}\bar{\boldsymbol{c}}\mathcal{D}\boldsymbol{c}\mathcal{D}\boldsymbol{\phi}e^{-\mathcal{S}}\label{part}\\
		  =& \sum_{km}\frac{1}{k!m!}\sum_{\nu_1\nu_1'\cdots\nu_k\nu_k'}\sum_{\sigma_1\sigma_1'\cdots\sigma_m\sigma_m'}\idotsint_{0}^{\beta} d\tau_1^c\tau_1^{c\prime}\cdots d\tau_k^c d\tau_k^{c\prime} d\tau_1^b\tau_2^b\cdots d\tau_{2m-1}^b d\tau_{2m}^b\nonumber \\
		  &\langle\bar{c}_{\nu_k}(\tau_k^c)c_{\nu_k'}(\tau_k^{c\prime})\cdots\bar{c}_{\nu_1}(\tau_1^c)c_{\nu_1'}(\tau_1^{c\prime})\phi_{\sigma_{2m}}(\tau_{2m}^b)\phi_{\sigma_{2m-1}}(\tau_{2m-1}^b)\cdots\phi_{\sigma_2}(\tau_2^b)\phi_{\sigma_1}(\tau_1^b)\rangle_{at}\nonumber\\
		  &\prod_{o=1}^{k}\Delta_{\nu_o\nu_o'}(\tau_o^c-\tau_o^{c\prime})\prod_{p,p'=1}^{2m}\prod_{p'\neq p}D_{\sigma_p\sigma_{p\prime}}(\tau_p^b-\tau_{p\prime}^b)\label{avg}\\
		  =&\sum_{km}\sum_{\boldsymbol{\nu}\boldsymbol{\sigma}}\int_{\tau_1^c >\cdots>\tau_k^c,\tau_1^{c\prime} >\cdots>\tau_k^{c\prime}} d\tau_1^c d\tau_1^{c\prime}\cdots d\tau_k^c d\tau_k^{c\prime}\int_{\tau_{1}^b >\tau_{2}^b>\cdots>\tau_{2m-1}^b>\tau_{2m}^b} d\tau_{1}^b \cdots d\tau_{2m-1}^b d\tau_{2m}^b\nonumber\nonumber\\
		  &\mathrm{Tr}\left[T_{\tau}e^{-\beta \mathcal{H}_{loc}}	
		  c^{\dagger}_{\nu_k}(\tau_k^c)c_{\nu_k'}(\tau_k^{c\prime})\cdots c^{\dagger}_{\nu_1}(\tau_1^c)c_{\nu_1'}(\tau_1^{c\prime})\phi_{\sigma_m}(\tau_m^b)\phi_{\sigma_m'}(\tau_m^{b\prime})\cdots\phi_{\sigma_1}(\tau_1^b)\phi_{\sigma_1'}(\tau_1^{b\prime})\right]\nonumber\\
		  &\mathrm{Det}\boldsymbol{\Delta}_k \prod_{p,p'=1}^{2m}\prod_{p'\neq p}D_{\sigma_p\sigma_{p\prime}}(\tau_p^b-\tau_{p\prime}^b).\label{ok}
	\end{align}	  
    Here, the average $\langle\cdots\rangle_{at}$ in Eq.~(\ref{avg}) is taken with respect to $\mathcal{S}_{at}$. In Eq.~(\ref{ok}), the determinant arises from a grouping procedure the same as that in the previous CT-HYB method. On the other hand, we do not apply such a grouping to the products of retarded interactions, but instead retain their original form. We recast the average in Eq.~(\ref{avg}) as a time-ordered average with respect to the local Hamiltonian $\mathcal{H}_{loc}$.  In Appendix~\ref{appA}, we discuss how the expansion factor is canceled by converting the full-range integrals to time-ordered integrals.
    
    \begin{figure}
    	\includegraphics[width=0.75\textwidth]{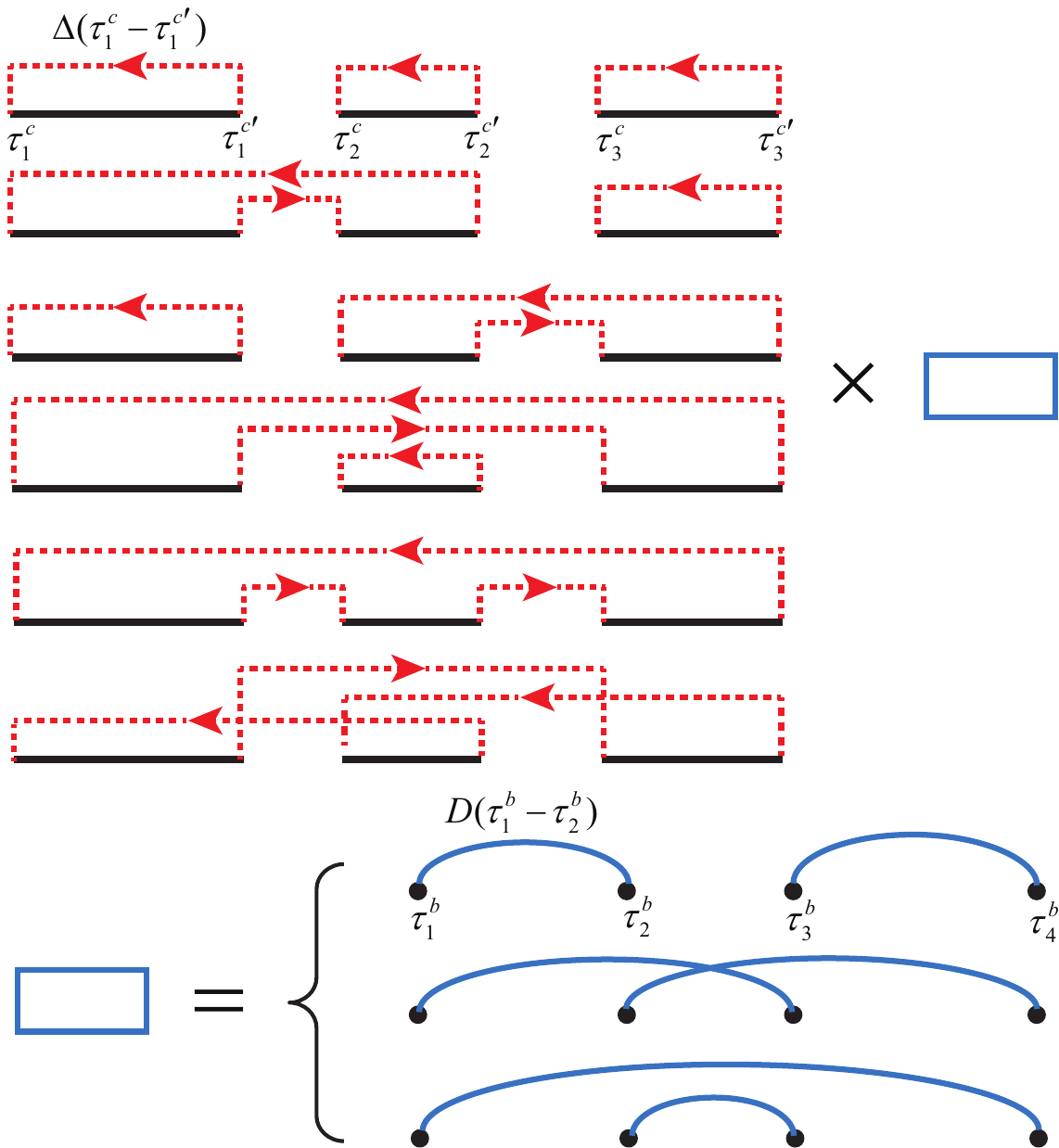}
    	\caption{A configuration contains both fermionic propagators (hybridization function) and bosonic propagators (retarded interactions). Grouping of these products of fermionic propagators gives rise to the determinant in Eq.~(\ref{ok}). We do not perform any grouping of the products of bosonic propagators, so the blue rectangle indicates one of the three products. \label{Fig1}}
    \end{figure}
	
	The above expansion implies that the configuration is extended to a hybrid form that includes both hybridization functions and retarded interactions. For example, we may diagrammatically represent a configuration with $k=3$ and $m=2$ as sketched in Fig.~\ref{Fig1}. The diagrams with red arrows denotes six possible ways to connect three pairs of time points with hybridization functions, and the combination of these diagrams leads to the determinant in Eq.~(\ref{ok}). The blue 
	rectangle represents one of the diagrams within the curly bracket. Here, we neglect variations in flavor. When these variations are considered, the number of states must be multiplied, since terms like $D_{\sigma\sigma'}(t_1^b-t_2^b)$ and $D_{\sigma'\sigma}(t_1^b-t_2^b)$ represent different propagators when $\sigma\neq\sigma'$.

	\subsection{Updates}
	For updates of the hybridization part, the update rules are identical to those in the previous CT-HYB algorithm. This is because, in the weight ratio between two configuration where the retarded interaction part remains unchanged, the product of bosonic propagators cancels out and has no effect. Therefore, we focus on the insertion and removal of a pair of bosonic operators. Labeling the current configuration as $\mathbf{x}$ and the configuration after the insertion as $\mathbf{y}$, the proposal probabilities for $\mathbf{x}\rightarrow\mathbf{y}$ and $\mathbf{y}\rightarrow\mathbf{x}$ are given by
	\begin{align}
		W_{\mathbf{xy}} &= \frac{d\tau^2}{\beta^2},\label{insert}\\
		W_{\mathbf{yx}} &= \frac{1}{m+1}.\label{rem}
	\end{align}
    Eq.~(\ref{insert}) presents the insertion of two operators in the (imaginary) time range $[0,\beta]$, with a mesh step of $d\tau$. Since we do not perform grouping operations on retarded interaction terms, a configuration of order $m$ in the retarded interaction part is simply a collection of $m$ pairs of bosonic operators. The denominator of $W_{\mathbf{yx}}$ is $m+1$, because their will be $m+1$ bosonic propagators after the insertion, and one of them is randomly selected for removal.

    The acceptance ratio of the insertion proposal can be derived from the detailed balance condition, which gives
	\begin{equation}
	R_{\mathbf{xy}} = \frac{p_{\mathbf{y}}W_{\mathbf{yx}}}{p_{\mathbf{x}}W_{\mathbf{xy}}}=\frac{\beta^2}{m+1}\frac{w_{loc}(\mathbf{y})\boldsymbol{D}(\mathbf{y})}{w_{loc}(\mathbf{x})\boldsymbol{D}(\mathbf{x})}.
	\end{equation}
    Here, $\boldsymbol{D}$ and $w_{loc}$ denote respectively the product of retarded interactions and the time-ordered average in Eq.~(\ref{ok}). The determinant for the hybridization functions is canceled. Since $\boldsymbol{D}(\mathbf{y})$ is equal to $\boldsymbol{D}(\mathbf{x})$ times the bosonic propagator corresponding to the inserted pair of operators, the ratio can be simplified to
    \begin{equation}
    	R_{\mathbf{xy}}=\frac{\beta^2}{m+1}\frac{w_{loc}(\mathbf{y})D_{\sigma\sigma'}(t_{\sigma}^b-t_{\sigma'}^b)}{w_{loc}(\mathbf{x})},
    \end{equation}
    where $t_{\sigma}^b$ and $t_{\sigma'}^b$ are two specific time points. The acceptance ratio for removing this pair is the inverse, i.e., $R_{\mathbf{yx}}=R_{\mathbf{xy}}^{-1}$.

    \subsection{Measurements}
    Since the Green's function measurement is the same as in the previous CT-HYB algorithm, our focus in on the correlation function measurement. A correlation function between two bosonic operators is conventionally defined as
    \begin{equation}
    	X_{\sigma\sigma'}(\tau-\tau') = -\langle\phi_{\sigma}(\tau)\phi_{\sigma'}(\tau')\rangle, 
    \end{equation}
    where the average is taken with respect to the total action. According to Eq.~(\ref{ret}) and Eq.~(\ref{part}), it is clear that the correlation function can be obtained by functional derivative with respect to the bosonic propagator. Due to the similarity between $\mathcal{S}_{ret}$ and $\mathcal{S}_{hyb}$, the derivation is similar to that of the Green's function measurement, for which one may refer to Ref.~\cite{Boehnke2011}. The formulation of correlation function measurement reads
    \begin{equation}
    	X_{\sigma\sigma'}(\tau) = \frac{1}{\beta}\left\langle\sum_{\rho\rho'}\delta_{\rho\sigma}\delta_{\rho'\sigma'} D_{\rho\rho'}^{-1}(\tau_{\rho}-\tau_{\rho'})\delta^+[\tau-(\tau_{\rho}-\tau_{\rho'})]\right\rangle_{MC},\label{dmeas}
    \end{equation}
    where $\langle\cdots\rangle_{MC}$ denotes the Monte Carlo average, and $\delta^+$ is the periodic delta function with a period of $\beta$. We call this scheme plain measurement to differentiate it with the combinatorial measurement proposed in the following.

    The symmetry of the correlation function can be leveraged to increase the number of measurements per sampling. The correlation function is essentially a bosonic Matsubara Green's function, so $X_{\sigma\sigma'}(\tau)=X_{\sigma'\sigma}(-\tau)=X_{\sigma'\sigma}(\beta-\tau)$. This symmetry can be enforced at the end as an overall average, or it can be implemented during each sampling. Specifically, whenever $X_{\sigma\sigma'}(\tau)$ is measured, we also measure $X_{\sigma'\sigma}(\beta-\tau)$. This approach effectively doubles the number of measurements. We implement this symmetry in all of our results.
    
    \begin{figure}
    	\includegraphics[width=0.7\textwidth]{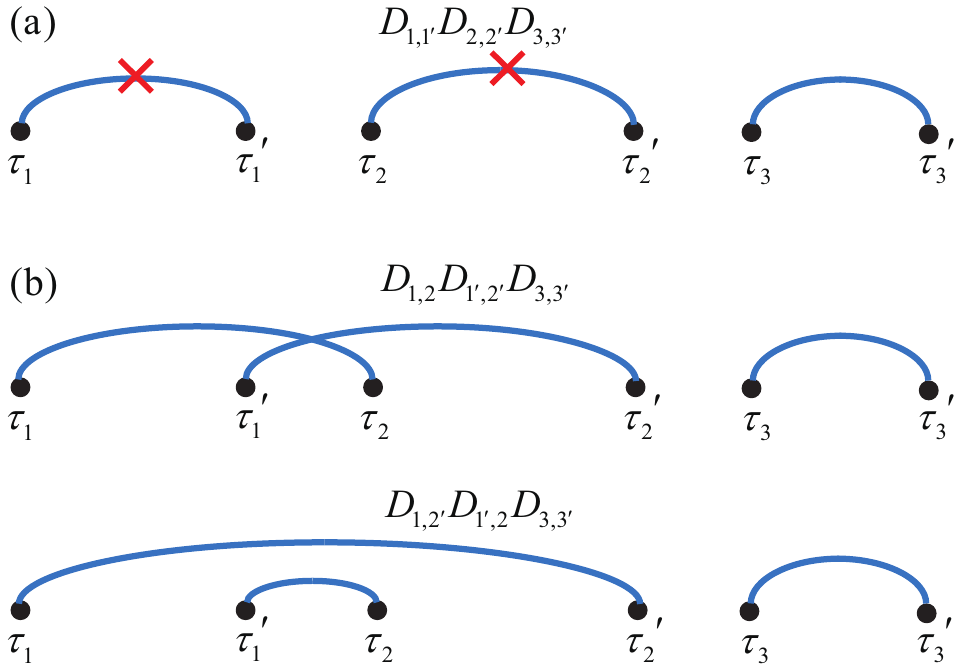}
    	\caption{(a) Cutting two bosonic operators results in four naked operators. (b) The naked operators can be repaired in two ways to produce two new configurations. This also means that the configuration in (a) can be obtained from the configurations in (b) via the cutting and repairing procedure.\label{Fig2}}
    \end{figure}

    For a sampled configuration of order $m$ in the retarded interaction part, it contains a product of $m$ $D_{\sigma\sigma}(\tau_{\sigma}-\tau_{\sigma'})$ propagators. According to Eq.~(\ref{dmeas}), only the correlation function corresponding to these propagators can be measured at this sampling. However, by cutting these propagators and repairing the bosonic operators, we can significantly increase the number of measurements. An example is illustrated in Fig.~\ref{Fig2}. After cutting the $D_{1,1'}$ and $D_{2,2'}$ propagators, there are two possible ways to relink the time points, which yield configurations $D_{1,2}D_{1',2'}D_{3,3'}$ and $D_{1,2'}D_{1',2}D_{3,3'}$, respectively. For simplicity, the time variable is not denoted explicitly here.

    In the probability form, Eq.~(\ref{dmeas}) amounts to
    \begin{equation}
    	X_{\sigma\sigma'} =\sum_{\mathcal{C}} p(\mathcal{C})\delta_{\rho\sigma}\delta_{\rho'\sigma'}D_{\sigma\sigma'}^{-1}=\sum_{\mathcal{C_{\sigma\sigma'}}}p(C_{\sigma\sigma'})D_{\sigma\sigma'}^{-1},
    \end{equation}
    where $\mathcal{C}$ denotes the whole configuration space, and $C_{\sigma\sigma'}$ represents the configurations containing the $D_{\sigma\sigma'}$ propagator. Let us further focus on the configuration $D_{1,2}D_{1',2'}D_{3,3'}$, which contributes to $X_{12}$ by the amount $p(12,1'2',33')D^{-1}_{12}$. Since the fermionic part (the determinant) and the time-ordered average $w_{loc}$ are intact in the cutting and repairing procedure, the weight ratio between configurations $D_{1,2}D_{1',2'}D_{3,3'}$ and $D_{1,1'}D_{2,2'}D_{3,3'}$ is
    \begin{equation}
    	\frac{p(12,1'2',33')}{p(11',22',33')} = \frac{D_{1,2}D_{1',2'}D_{3,3'}}{D_{1,1'}D_{2,2'}D_{3,3'}}=\frac{D_{1,2}D_{1',2'}}{D_{1,1'}D_{2,2'}}.
    \end{equation}
    Accordingly, we have
    \begin{equation}
    	p(12,1'2',33')D^{-1}_{1,2} = p(11',22',33')\frac{D_{1',2'}}{D_{1,1'}D_{2,2'}}.
    \end{equation}
    The left hand side is exactly the measured value of $X_{12}$ from configuration $D_{1,2}D_{1',2'}D_{3,3'}$, and the right hand side is an measurement that can be performed based on the sampling of configuration $D_{1,1'}D_{2,2'}D_{3,3'}$. In other words, when configuration $D_{1,2}D_{1',2'}D_{3,3'}$ is sampled, $X_{12}$ can be measured as $D_{1',2'}/(D_{1,1'}D_{2,2'})$.

    The combinatorial measurement of correlation function can be generally formulated as
    \begin{align}
       X_{\sigma\sigma'}(\tau) & = \frac{1}{\beta}\left\langle\sum_{\rho\rho'}\delta_{\rho\sigma}\delta_{\rho'\sigma'}\frac{1}{O(\mathbf{D})} D_{\rho\rho'}^{-1}(\tau_{\rho}-\tau_{\rho'})\delta^+[\tau-(\tau_{\rho}-\tau_{\rho'})]\right\rangle_{MC}\nonumber\\
       & +\frac{1}{\beta}\left\langle\sum_{\eta\eta'\rho\rho'}\delta_{\rho\sigma}\delta_{\rho'\sigma'}\frac{1}{2O(\mathbf{D})} \frac{D_{\eta\eta'}(\tau_{\eta}-\tau_{\eta'})}{D_{\rho\eta}(\tau_{\rho}-\tau_{\eta})D_{\rho'\eta'}(\tau_{\rho'}-\tau_{\eta'})}\delta^+[\tau-(\tau_{\rho}-\tau_{\rho'})]\right\rangle_{MC},\label{cmear}
    \end{align}
    where $O(\mathbf{D})$ denotes order of the product of bosonic propagators. The first line is the plain measurement, and the second line represents the combinatorial scheme. To see why we weigh the measurements with $O(\mathbf{D})$, suppose we are measuring $X_{11'}$. It can be directly measure from configuration $D_{1,1'}D_{2,2'}D_{3,3'}$. We can also measure it from configurations that differ by a cutting and repairing procedure. In this example, there is two ways (propagators $1,2$ or propagators $1,3$) of selecting a propagator pair for cutting, and the four unpaired operators can be repaired in two ways [cf. Fig.~\ref{Fig2}(b)]. By reversing the logic, this means that the cutting and repairing procedure can offer $2\times2$ equivalent measurements of $X_{11'}$. In this simple example, $O(\mathbf{D})=3$, so the effective number of measurements is $1/3+(2\times2)/(2\times3)=1$. Namely, we do one effective measurement by averaging over $5$ measurements. For $O(\mathbf{D})=m$, we have $1/m+[(m-1)\times2]/(2\times m)=1$.

    In the implementation, when a certain configuration is sampled, besides the plain measurement [first line of Eq.~(\ref{cmear})], the cutting and repairing procedure is applied to any two propagators in a configuration. This way, the correlation function between any two time points are measured, so the efficiency is comparable to the Green's function measurement based on the determinant in Eq.~(\ref{ok}). One could select three or more propagators for the cutting and repairing procedure to further increase the efficiency, whereas this would supposedly complicate the algorithm.

	\section{Test case\label{III}}
	\subsection{Model}
	We apply the proposed algorithm to an impurity model with two orbitals, where Coulomb interaction and inter-orbital exchange interactions are present, and the electrons couple to a phonon mode via Holstein-type interactions. The local Hamiltonian is given by
	\begin{equation}
		\mathcal{H}_{loc} = \sum_{i=1,2}Un_{i\uparrow}n_{i\downarrow} + \sum_{\sigma\sigma'=\uparrow\downarrow}U'n_{1\sigma}n_{2\sigma'} - J\mathbf{s_1}\cdot\mathbf{s_2} + \sum_{i\sigma}\mu n_{i\sigma}.
	\end{equation}
	After integrating out the bath and phonon degrees of freedom, respectively, the hybridization functions and retarded interactions are described by~\cite{CTMC_RMP2006,Werner2010,Assaad2007}
	\begin{align}
		&\mathcal{S}_{hyb} = -V^*V\sum_{i\sigma}\iint_{0}^{\beta}d\tau d\tau'\bar{c}_{i\sigma}(\tau)\frac{e^{-\epsilon(\tau-\tau'-\beta)}}{e^{\epsilon\beta}+1}c_{i\sigma}(\tau'),\\
		&\mathcal{S}_{ret} = -\lambda^2\sum_{ij}\sum_{\sigma\sigma'}\iint_{0}^{\beta}d\tau d\tau'n_{i\sigma}(\tau)\frac{\cosh[(\tau-\tau'-\beta/2)\omega_{\lambda}]}{\sinh(\beta\omega_{\lambda}/2)}n_{j\sigma'}(\tau'),
	\end{align} 
	Here $V$ and $\lambda$ represent the coupling coefficients for coupling to the bath and the Holstein electron-phonon interaction, where energies of the bath fermion and the phonon mode are $\epsilon$ and $\omega_{\lambda}$, respectively.

	We set the chemical potential as $\mu=1$, which defines the energy unit of the systems. The other parameters are set as follows: $\beta=10$ ($T=0.1$), $U=2$, $J=0.2$, $U'=U-2J$~\cite{Winter2016}, $V=0.6$, $\lambda=0.19$, $\omega_{\label{key}}=0.1$, $\epsilon=0.02$. Our numerical implementation is based on the CTHYB implementation~\cite{Seth2016} within the TRIQS framework~\cite{Parcollet2015}, with substantial extensions to include retarded interactions and the measurements of correlation functions. With these settings, the average orders of both the hybridization function and retarded interaction are about $10$. An inserted pair of bosonic operators are made by randomly selecting two of the four density operators, so $W_{\mathbf{xy}}$ in Eq.~(\ref{insert}) should be further divided by $16$. We also note that in both calculations with and without the retarded interaction, the average sign is approximately $1$. In other words, these is no sign problem for this model, and the inclusion of the retarded interaction does no harm in this case.

	\subsection{Results}
\begin{figure}
		\includegraphics[width=0.8\textwidth]{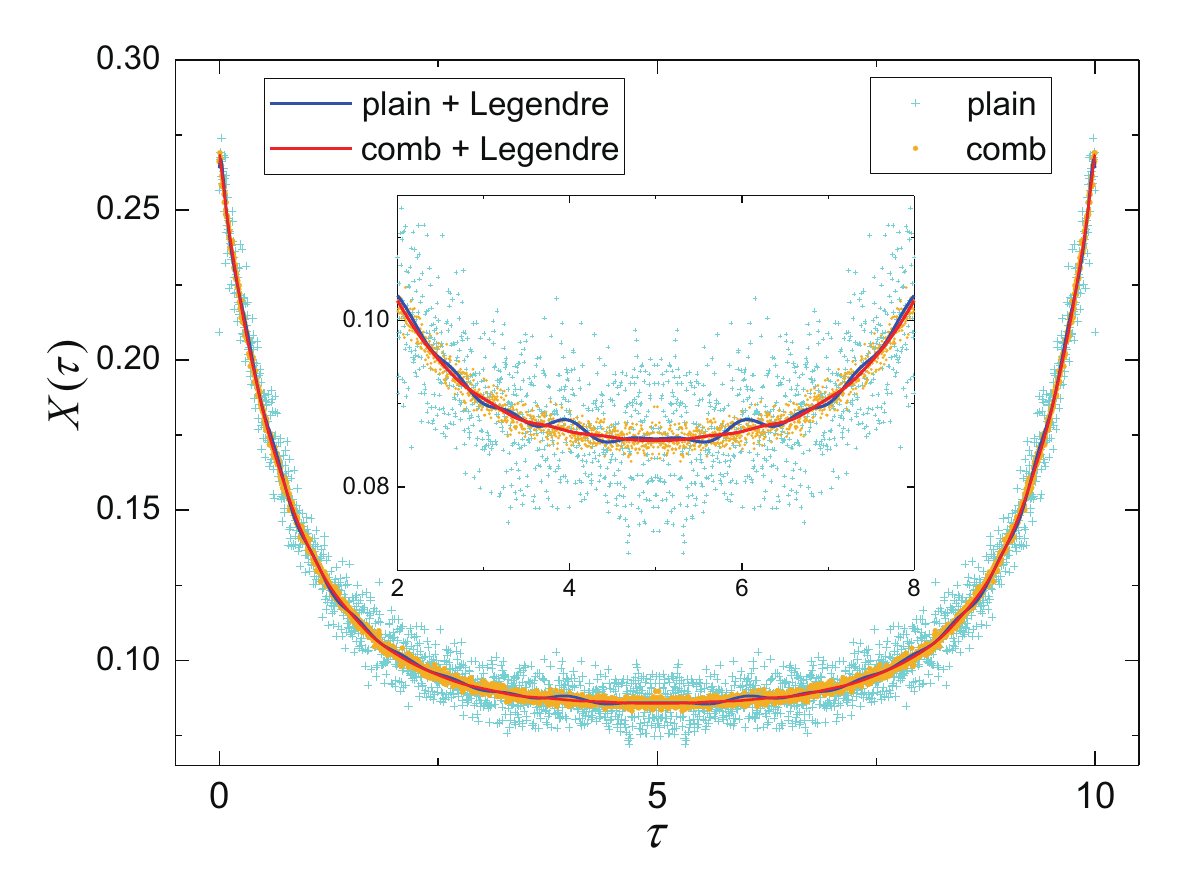}
		\caption{The combinatorial measurement (yellow dots) yields much less noisy results than the plain measurement (cyan cross). The results transformed from the Legendre representation are significantly denoised and becomes smooth curves. The zoom-in in the inset suggests that the combinatorial measurement regularizes the result. \label{Fig3}}
	\end{figure}

	We present the correlation function $-\langle n_{1\uparrow}(\tau)n_{1\uparrow}(0)\rangle$ below. The scatter plots in Fig.~\ref{Fig3} demonstrate the correlation function measured using the plain measurement (Eq.~\ref{dmeas}) and the combinatorial measurement (Eq.~\ref{cmear}). It is clear that the combinatorial scheme leads to much less noisy results. For an configuration of order $m$ in the retarded interaction part, the plain scheme simply performs $m$ measurements, while the combinatorial scheme conducts $m+2m(m-1)$ measurements. In other words, the number of measurements is approximately increased by a factor of $2m$. Since the average order is $\langle m\rangle\approx10$ here, we can conclude that the measurement efficiency is enhanced by about $20$ times.
	
	Since the formulation of the correlation function measurement is similar to that of the Green's function measurements, the Legendre representation proposed in Ref.~\cite{Boehnke2011} is directly applicable to both the direct and combinatorial schemes. To demonstrate the denoising effect, we transform the Legendre representation back to the imaginary-time correlation. As can be seen, the results become smooth curves. The inset clearly shows that the result enhanced by the combinatorial scheme is more regular. The order of the Legendre polynomials is set to $l=50$.

\begin{figure}
		\includegraphics[width=0.8\textwidth]{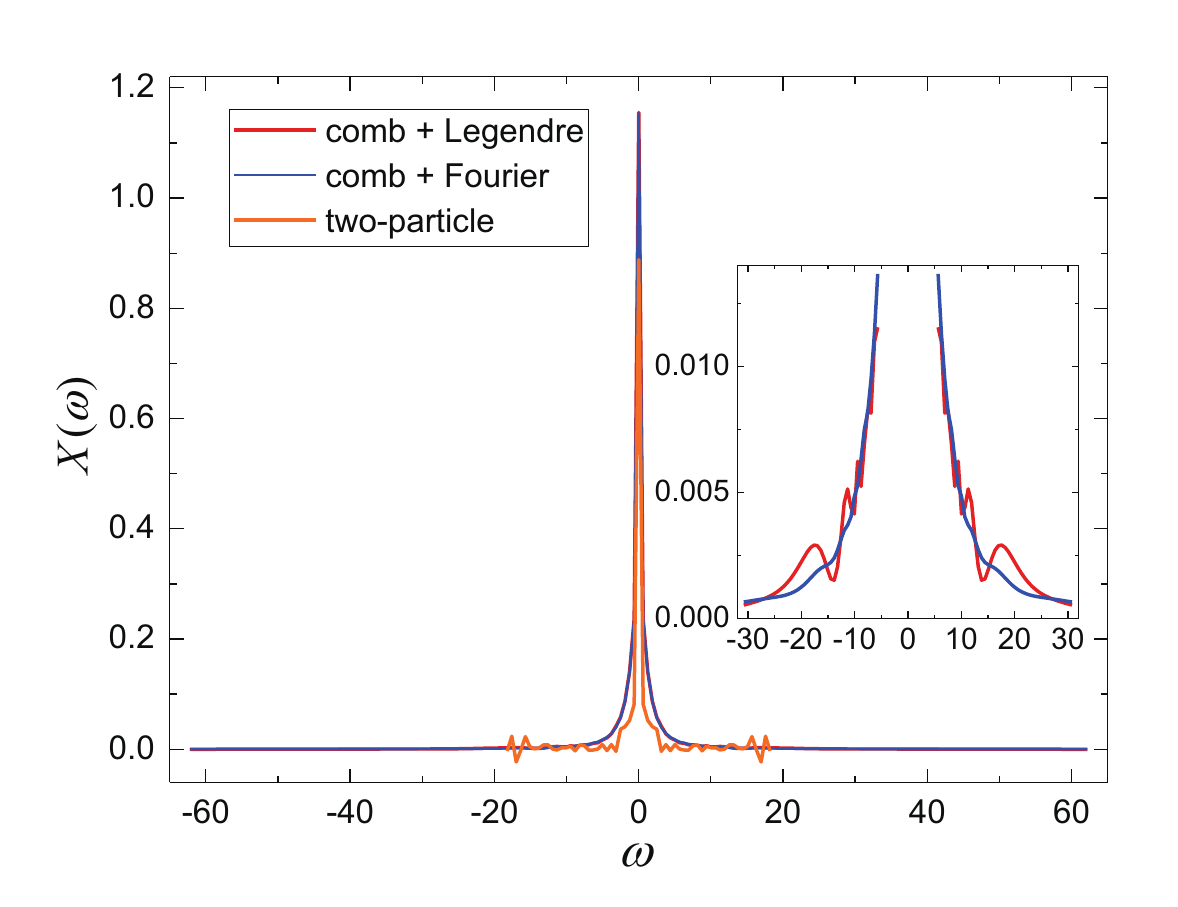}
		\caption{When transformed to the imaginary-frequency domain, the correlation function measured directly in the imaginary time domain is more regular than the that using Legendre representation. The correlation function derived from the corresponding two-particle Green's function is irregular, which can be ascribed to an inconsistency of the measurement procedure for impurities of finite bath~\cite{Hausoel2022}.\label{Fig4}}
	\end{figure}

	In Fig.~\ref{Fig4}, we transform the combinatorially measured results into the frequency domain, where the imaginary-time correlation function undergoes a Fourier transform, and the Legendre representation is directly converted via the Legendre-to-Matsubara transform~\cite{Boehnke2011}. Notably, although the correlation measured directly in the imaginary time domain is much noisier, the corresponding imaginary-frequency function is smoother (see the zoom-in). It appears that the Legendre representation causes some loss of information.

	We also show the correlation function derived from two-particle Green's functions in the mixed Legendre-Fourier basis~\cite{Boehnke2011}, where the order of the Legendre polynomials is set to $l=20$. The irregular results suggest that the measurement of the two-particle Green's function fails to yield consistent results. This can be attributed to the missing configurations due to the Pauli exclusion principle, when the impurity electrons have more flavors than the auxiliary base electrons (see Chapter 4 of Ref.~\cite{Hausoel2022}). Technically, this inconsistency is related to the vanishing of the determinant in Eq.~(\ref{ok}). Since the retarded interaction part does not affect the manipulation of the determinant, the correlation function measurement based on functional derivative with respect to the bosonic propagator is free from this inconsistency.

	\section{Conclusions and outlook}
	In summary, we propose a general hybridization expansion approach for impurity models that include both hybridization functions and retarded interactions. By explicitly incorporating retarded interactions into the expansion, the density-density and spin-spin correlation functions can be measured via functional derivative with respect to the bosonic propagator. The measurement efficiency is further enhanced by a combinatorial scheme based on propagator cutting and operator repairing. This approach allows for the measurement of correlation functions between any two involved time points, sizably increasing the number of measurements per sampling. Additionally, compact representation techniques for Green's functions can be easily incorporated into the measurement procedure, which significantly reduces noise in the results.

	Compared to the previous approach to handle the retarded interaction, which is limited to the density-density interactions, the proposed CT-HYB method impose no such restrictions. In addition to its application in DMFT investigation of electron-phonon interaction, a major potential usage of this algorithm is as a solver for the effective impurity model in the EDMFT and the dual Boson method, particularly when inter-site exchange interactions are present. Since the hybridization functions and retarded interactions are treated on equal footing, their corresponding propagators can both be determined self-consistently. Application of this extended CT-HYB solver to these problems may yield more reliable results and enhance our understanding of such systems.

	In our calculations, we find that changing the coupling strengths ($V$ and $\lambda$) may result in a low average order for either the hybridization part or the retarded interaction part. Physically, this occurs because the partition function is an exponential function of energy, making the probability distribution sensitive to interaction strength. For low other regimes, worm sampling~\cite{Gunacker2015} is a general method to obtain reliable measurements. It is also worth exploring whether the inchworm quasi Monte Carlo method~\cite{Eidelstein2020,Strand2024} can improve  measurement precision and mitigate potential sign problem in the context of the extended CT-HYB solver. By cutting the bosonic propagators and relinking the compositional fermionic operator with the hybridization functions, a retarded interaction term can be transformed into hybridization terms. This form transformation may imply new update strategies and measurement procedures.

	\begin{acknowledgments}
	L.G. is grateful to A. Hausoel, I. Krivenko, H. Hafermann and N. Wentzell for valuable explanations on DMFT and code implementation. L.G. thanks the support from the Sichuan Normal University.  R.W. acknowledges the support from DOE-BES (Grant No. DE-FG02-05ER46237). G.Z. acknowledges the support from the National Natural Science Foundation of China (Grant Nos. 12474122, 52171188, 51771127 and 52111530143) and the Central Government Funds of Guiding Local Scientific and Technological Development for Sichuan Province (No. 2021ZYD0025). 
	\end{acknowledgments}

\appendix
\section{Cancellation of expansion factors\label{appA}}
Instead of expansion with respect to the hybridization functions and retarded interactions, here we return to the point where the bath operators are paired to form the propagators. We first discuss the hybridization function. To form $k$ propagators, the number of creation and annihilation operators should both be $k$, i.e., expansion terms of order $2k$. The number of such terms are $C(2k,k)$. The next step is to assign $k$ (imaginary) time variables to the creation operators and another set of $k$ time variables to the annihilation operators. When the time integral is from $0$ to $\beta$, any arrangement of time variables will yield a specific configuration. Therefore, the number of ways to assign the time variables is $k!k!$. In other words, in the form of full-range integrals, the number of occurrence of a specific configuration is $C(2k,k)k!k!$, which exactly cancels the expansion factor $1/(2k)!$.

When two types of bosonic operators of distinct flavors present and their numbers are identical, the argument remains the same. We now further discuss the pairing of bosonic operators of the same flavor. In this case, we need to assign $2k$ time variables to $2k$ identical operators, the number of ways to do this is exactly $(2k)!$. It is important to note that the cancellation of the expansion factor has nothing to do with grouping. To get the determinant in Eq.~(\ref{ok}), one simply collects terms having the same time-ordered average, and no additional operations are required.

When two types of bosonic operators present with unequal numbers ($n$ and $2k-n$, respectively), the number of such terms is $C(2k,n)$. There are $n!$ ways to assign $n$ time variables to the $n$ identical operators. Selecting $n$ time variables from $2k-n$ candidates for the operators to be paired with the $n$ operators lead to a factor of $C(2k-n,n)$, and there are $n!$ ways to do the paring. The remaining $2k-2n$ time variables can be assign to the remaining $2k-n$ operators in (2k-2n)! ways. Since $C(2k,n)n!C(2k-n,n)n!(2k-2n)!=(2k)!$, the expansion factor is canceled. More complicated cases can be reasoned similarly.

From the perspective of time-variable assignment, it is easy to understand why the expansion factor is $1$ in the form of time-ordered integrals. Given a time-ordering constraint, such as $t_1>t_2>\cdots>t_{2n}$, there can be only one valid assignment scheme for a specific configuration, where the operators locate at different time points. In other words, a specific configuration occurs only once in the expansion with time-ordered integrals.

\section{Results of an off-diagonal correlation}

\begin{figure}[b]
	\includegraphics[width=0.7\textwidth]{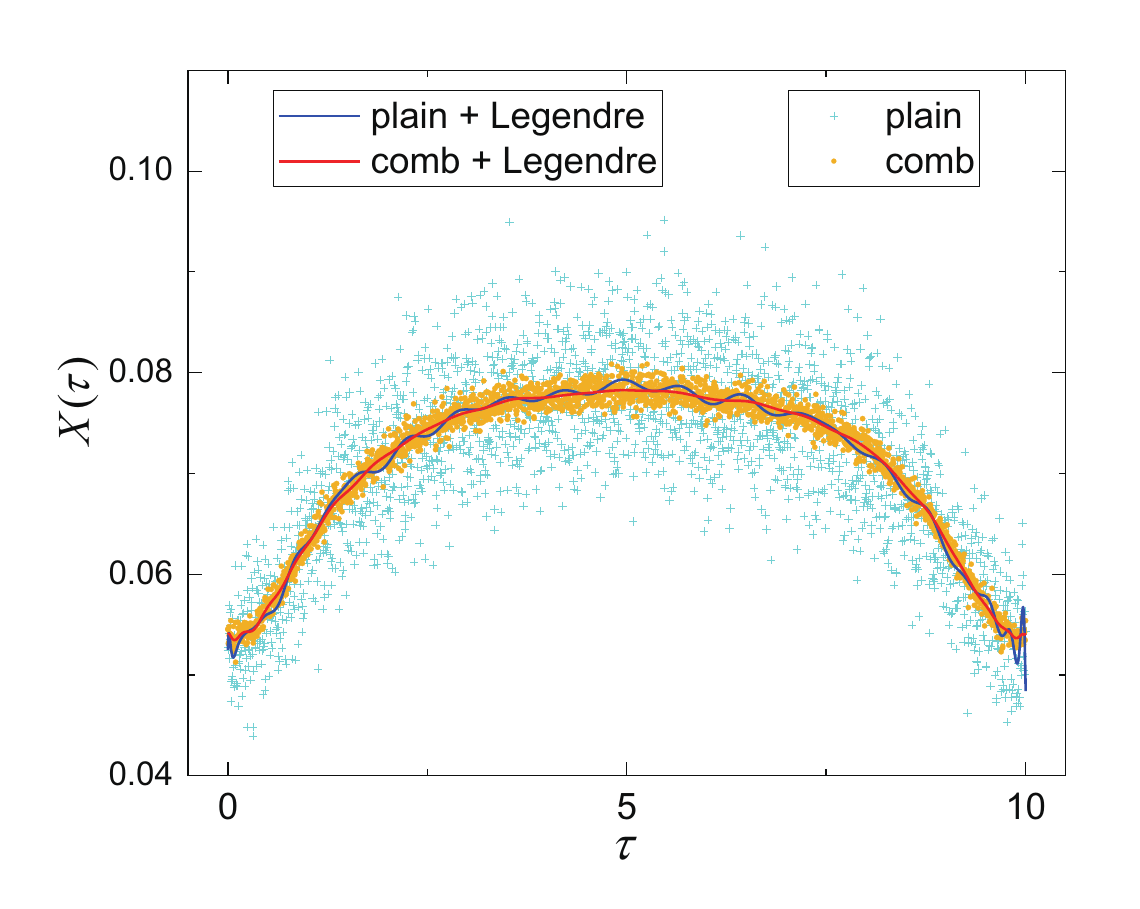}
	\caption{Results of correlation $-\langle n_{1\uparrow}(\tau)n_{2\downarrow}(0)\rangle$. Similar to the main text results, it shows that the combinatorial measurement scheme has denoising and regularizing effects, and the Legendre representation can significantly denoise the result. \label{Fig5}}
\end{figure}

In Fig.~\ref{Fig5}, we present the correlation function $-\langle n_{1\uparrow}(\tau)n_{2\downarrow}(0)\rangle$. The utilities of the combinatorial measurement and Legendre representation are the same as those discussed in the main text.

\bibliography{dmft_ref.bib}

\end{document}